
\magnification=1200
\baselineskip=20pt
\tolerance=10000
\line{\hfill UR-1413}
\line{\hfill ER-40685-860}
\bigskip
\centerline{Perturbation Theory and the Aharonov-Bohm Effect}
\medskip
\centerline{by}
\centerline{C. R. Hagen}
\centerline{Department of Physics and Astromony}
\centerline{University of Rochester}
\centerline{Rochester, NY 14627}
\bigskip
\noindent Abstract

The perturbation theory expansion of the Aharonov-Bohm scattering
amplitude has previously been studied in the context of quantum mechanics
for spin zero and spin-1/2 particles as well in Galilean covariant field
theory.  This problem is reconsidered in the framework of the model in
which the flux line is considered to have a finite radius which is shrunk
to zero at the end of the calculation.  General agreement with earlier
results is obtained but with the advantage of a treatment which unifies
all the various subcases.

\vfil\eject
\noindent I.  Introduction

The Aharonov-Bohm effect [1] is generally considered to be among the more
intriguing predictions of quantum mechanics.  One particular reason for
this is the fact that it requires that the potential itself be viewed as
having a physical significance which transcends its role as a mere
mathematical construct for the calculation of a classical force.
However, there are also some mathematical aspects of this phenomenon
which are quite intriguing.  Not the least of these is the study of
the scattering amplitude in perturbation theory as was pointed out [2]
some years ago.

In order to understand the nature of the difficulty it is convenient to
refer to the wave equation for the partial wave $f_m(r)$.  It can be
written as
$$\left[ {1\over r} {\partial\over \partial r} r {\partial\over \partial
r} + k^2 - {(m+\alpha)^2\over r^2}\right] f_m(r)=0$$
where $\alpha$ is the flux parameter and $k^2=2ME$ with $M$ the particle
mass and $E$ its energy.  If one applies standard techniques and
invokes the condition $f_m(0)=0$, it is found that the phase shifts can be
written as
$$\delta_m = {-\pi\over 2} | m+\alpha | + {\pi\over 2} |m|$$
so that for small $\alpha$
$$\delta_0 = -|\alpha |\pi/2\>.$$
This result is somewhat disturbing since it implies that the $m=0$
contribution to the scattering amplitude is of order $\alpha$ even
though the $m=0$ potential is proportional to the square of the flux parameter.

This aspect was examined in some detail by Aharonov {\it et al.} [3] in the
framework of a model in which the solenoid was made impenetrable and of
finite radius $R$.  Although in the limit $R\to 0$ the exact solution for
finite $R$ was found to
 become identical to the usual Aharonov-Bohm (AB) amplitude, to
any finite order in $\alpha$ the solution took the form of a complicated
expansion in powers of $\alpha \ell n (kR/2)$.  The results of ref. 3
were obtained by retaining only that part of the $m=0$ wave function
proportional to $J_\alpha(kr)$ and discarding terms involving the
solution $N_\alpha(kr)$ which is singular at the origin.  The impropriety
of ignoring that part of the solution has been remarked upon by Nagel.
In fact the inclusion of such terms affects the quantitative, though not
the qualitative, results of ref. 3.

In fact one can anticipate on the basis of fairly general considerations
that any perturbative expansion must experience some difficulty.  To this
end one refers to the (slightly corrected) form of the AB amplitude given
in [5].  There one has
$$f_{\rm AB}(\phi) = - \left({i\over 2\pi k}\right)^{1/2} {\sin\pi\alpha
e^{-iN(\phi-\pi)}\over \cos \phi/2} e^{-i\phi/2} \eqno (1)$$
where the incident wave is assumed to be from the right and $N$ is the
largest integer in $\alpha$, i.e.,
$$\alpha = N + \beta$$
with
$$0 \leq \beta < 1\>.$$
For small $\alpha$ one readily obtains from (1) the result
$$f_{\rm AB}(\phi) \to \alpha \left({i\pi\over 2k}\right)^{1/2}\left(i\tan
\phi/2 -\epsilon(\alpha)\right)\eqno(2)$$
where
$$\epsilon(\alpha) \equiv \alpha/|\alpha | \>.$$
Evidently, the second term in (2) (which arises from the $m=0$ wave
contribution to the amplitude) is nonanalytic in $\alpha$, a fact which
can be expected to complicate any perturbative expansion of the amplitude
in powers of $\alpha$.

In order to display most effectively the problems associated with a
perturbative approach to the Aharonov-Bohm effect, it is crucial to
use a unified model which allows one to consider simultaneously both the
spinless and spin-1/2 cases.  The details of such a model are presented
in the following section.  Since it is known [6] that the lowest order
Born approximation works well in the spinor case, it is important to
establish that the model agrees with that lowest order result and also
that it accounts for the difficulties of the spinless case noted in refs.
1-4.  In sec. 3 this program is carried out. It is shown there that the
perturbative expansion works to all orders in $\alpha$ in the spin-1/2
case for both spin orientations and also that there exist
logarithmic singularities in each order for the case that
there is no spin degree of freedom.  In the Conclusion some general
observations are presented and contact made with results which have been
obtained in the context of Galilean field theory.

\noindent II.  The Finite Radius Flux Tube Model

It has already been noted that the perturbative treatment of the AB
effect given in ref. 3 was based on an impenetrable solenoid of radius $R$.
Thus
the wave function which solves Eq. (1) is a combination of the regular
Bessel function $J_{|m+\alpha |}(kr)$ and the irregular one
$N_{|m+\alpha|}(kr)$. Specifically
$$f_m(r) \sim \left[N_{|m+\alpha |}(kR)J_{|m+\alpha |}(kr)-
J_{|m+\alpha |}(kR)N_{|m+\alpha |}(kr)\right]\>.$$
In the case of AB scattering for spin-1/2 particles, however, it is well
known [7] that for a flux tube of zero radius the wave equation (1) must
be replaced by
$$\left[ {1\over r}{\partial\over\partial r}r{\partial\over \partial r}+k^2
-{(m+\alpha)^2\over r^2} - \alpha s {1\over r}\delta
(r)\right]f_m(r)=0\eqno(3)$$
where $s \pm 1$ is the spin projection parameter.  The delta function
term evidently describes the interaction of the particle's magnetic moment
with the magnetic field of the flux tube.  If one were to adopt the
impenetrable solenoid model of ref. 3, it is clear that the magnetic
moment term would be rendered ineffective and there could be no spin
effect whatever.  In view of the desire expressed here to develop a
unified approach which will accommodate both spin zero and spin-1/2, a
rather different model (or regularization) is therefore required.

Such a model has in fact been presented in the context of obtaining the
solution of the spin-1/2 AB scattering amplitude [7].  It consists of
replacing the zero radius flux tube by one of radius $R$, with the
additional condition that the magnetic field be confined to the surface
of the tube [8].In this case Eq. (3) is replaced by
$$\left[{1\over r}{\partial\over \partial r} r {\partial\over \partial r}
+ k^2 - {1\over r^2} [m+\alpha\theta (r-R)]^2 -\alpha s {1\over R}\delta
(r-R)\right] f_m(r)=0\eqno(4)$$
where $\theta (x)$ is the usual step function
$$\theta(x) = {1\over 2} [1+\epsilon (x)].$$
Since the ultimate interest here is in a perturbation expansion, there is
no need to consider any partial wave except $m=0$, the only case in which
a perturbative failure can occur.  Thus one writes the solution of (4) as
$$f_0(r) = AJ_\alpha (kr) + BN_\alpha (kr)\eqno(5)$$
for $r>R$ and
$$f_0(r) = CJ_0(kr)$$
for $r<R$.  It is to be understood that even though $\alpha$ can be of
either sign, it is always to be taken to mean $|\alpha |$ when used to
denote the order of a Bessel function.  Also worth mentioning is the fact
that although the calculations of ref. 7 based on this model used the
functions $J_{\pm |m + \alpha |}(kr)$ to describe the $r>R$ solution,
these are an inappropriate set for a perturbative analysis.  The latter
must maintain consistency for $\alpha \to 0$, a requirement clearly
violated by the functions $J_{\pm\alpha}(kr)$ which become identical
in that limit.\break
\parskip=0pt plus 1pt
\indent From (4) one obtains the continuity relations
$$\eqalign{f_0(R-\epsilon) &= f_0 (R+\epsilon)\cr
\noalign{\vskip3pt}
R{d\over dR}f_0 \bigg\vert ^{R+\epsilon}_{R-\epsilon} &= \alpha s f_0(R)\cr}$$
which are readily seen to imply
$$CJ_0(kR) = AJ_\alpha (kR) + BN_\alpha(kR)$$
$$\left( R{\partial\over \partial R} + \alpha s\right) CJ_0(kR) = AR
{\partial\over \partial R} J_\alpha (kR) + BR {\partial\over \partial R}
N_\alpha (kR).$$
This yields for the ratio of irregular to regular part of the wave
function the result
$$B/A = {J_0R{\partial\over \partial R}J_\alpha - J_\alpha
(R{\partial\over \partial R} + \alpha s)J_0\over
N_\alpha (R {\partial\over\partial R} + \alpha s)J_0 - J_0
R{\partial\over \partial R} N_\alpha}\>.\eqno(6)$$
Since one is interested in the $R \to 0$ limit, it is permissible to drop
all terms which differ from the leading term by one or more powers of
$kR$.  Thus (6) becomes
$$B/A = {(R{\partial\over \partial R}-\alpha s)J_\alpha\over
(-R{\partial\over \partial R} + \alpha s)N_\alpha}\>.\eqno(7)$$

In order to make contact with perturbation theory one notes that the
$m=0$ contribution to the wave function for $r \to \infty$ is of the form
$$f_0(r) \sim J_0(kr) + f_0 e^{ikr}/r^{1/2}\>.\eqno(8)$$
with the first term in (8) arising from the incident plane wave.  Upon
comparison of (5) and (8) one finds that
$$Ae^{i|\alpha|\pi/2} + Be^{-i|\alpha|\pi/2} = 1.$$
Use is  now made of the fact that $f_0(r)$ can be written as
$$f_0(r) = J_0(kr) + \int ^\infty_0 r^\prime dr^\prime g(r,r^\prime)
\left[ {\alpha^2\over r^{\prime 2}}\theta (r^\prime-R) + {\alpha s\over R}
\delta (r^\prime -R)\right] f_0(r^\prime)$$
where  the Green's function $g(r,r^\prime)$ has the form
$$g(r,r^\prime) = -{i\pi\over 2} J_0 (kr_<)H^{(1)}_0(kr_>)$$
with
$$H_0^{(1)}(x) \equiv J_0(x) + iN_0(x).$$
This allows the $m=0$ scattering amplitude to be identified as
$$\eqalign{f_0 &= -i\left({\pi\over 2ik}\right)^{1/2} {e^{-i\pi
|\alpha|/2}\over 1 + i(B/A)} \biggl\{\alpha s J_0 (kR) \biggl[J_\alpha (kR)
+ { B\over  A} N_\alpha (kR)\biggr]\cr
\quad&+ \alpha^2 \int^\infty_R {dr\over r} J_0 (kr)\left[ J_\alpha (kr) +
{ B \over  A} N_\alpha (kr)\right]\biggr \}\cr}\eqno(9)$$

The result (9) together with (7) allows one in  appropriate limits to
establish the properties of the scattering amplitude as a function of
$\alpha$ and $R$ for both the spin zero and spin-1/2 cases.  For the
latter one considers both $s = +1$ and $s= -1$ while in the spinless case
one merely has to set the spin parameter $s$ equal to zero.  The results
in these various cases are presented in the following section.

\noindent III.  The Perturbative Expansion

Since the spin-1/2 case is expected to have the fewest complications in
perturbation theory, it is natural to begin with that example.  One also
anticipates on the basis of ref. 7 that the repulsive (i.e., $\alpha s
>0$) delta function will be particularly simple.  Indeed, from (7) one
finds that in that situation $B/A$ is zero to the required order in $kR$
and that (9) thus reduces to
$$f_0 = -i \left({\pi\over 2ik}\right)^{1/2} e^{-i|\alpha|\pi/2}
\left\{|\alpha|J_\alpha(kR) + \alpha^2 \int^\infty_R {dr\over r} J_0
(kr) J_\alpha(kr)\right\}\eqno(10)$$
which by elementary means becomes
$$f_0 = -i\left({\pi\over 2ik}\right)^{1/2} e^{-i|\alpha|\pi/2}
{2\over \pi}\sin{|\alpha|\pi\over 2}$$
or
$$f_0 = \left({1\over 2\pi k i}\right)^{1/2}
(e^{-i|\alpha|\pi}-1)\>.\eqno(11)$$
The most significant aspect of this result from the present perspective
is that the dependence upon $R$ has vanished since the leading term in
$kR$ goes as a positive power of $kR$ greater than one.  Had
that power instead been proportional to $\alpha$ an infinite series in
$\alpha \ell n kR$ would have been encountered in perturbation theory.
In fact each of the two terms in (10) contains a term proportional to
($kR)^\alpha$ and it is only as a consequence of a delicate cancellation
between the spin term and the non-spin term that the $R$ independent
result (11) is obtained.

Somewhat more intricate is the attractive case $\alpha s < 0$.  In this
circumstance application of (7) yields the result
$$ B/A = - \tan \pi |\alpha| .\eqno(12)$$
Insertion of (12) into (10) yields upon evaluation of the integrals
$$\eqalign{f_0 &= -i \left({\pi\over 2ik}\right)^{1/2} e^{i|\alpha|\pi/2}
\biggl\{-2|\alpha|J_\alpha (kR) + |\alpha|\tan \pi|\alpha|N_\alpha(kR) +
{2\over \pi}\sin {\pi|\alpha|\over 2}\cr
&\quad -\tan\pi|\alpha|\left[ {2\over \pi}\cos {\pi\alpha\over 2} - R
{\partial\over \partial R} N_\alpha (kR)\right]\biggr\}\cr}$$
which is readily reduced to
$$f_0 = \left({1\over 2\pi ki}\right)^{1/2} \left(e^{i\pi|\alpha|}-1
\right)\>.\eqno(13)$$
Again, a significant cancellation of spin dependent and spin independent
terms has resulted in a disappearance of $R$ dependent terms to the order
required.

The results (11) and (13) can be combined in the single expression
$$f_0^{s=1/2} = \left({1\over 2\pi ki}\right)^{1/2}
\left[ e^{-i\alpha\epsilon(s)}-1\right]\>$$
It confirms the lowest order Born approximation result of ref. 6 by
virtue of its independence of the flux tube radius as well as by its
analyticity in the flux parameter.  Clearly, perturbation theory works to
all orders for spin-1/2.

These results are significantly changed in the $s=0$ case.  Application
of (7) and (9) yields.
$$f_0^{s=0} = -i \left({2\over \pi ik}\right)^{1/2} e^{-i|\alpha|\pi/2}
 {1 + {B\over A} \cot {|\alpha|\pi\over 2}\over
1 + {iB\over A}}\sin {\pi|\alpha|\over 2}\eqno(14)$$
where
$$B/A = - {R {\partial\over\partial R} J_\alpha (kR)\over N {\partial
\over\partial R}
N_\alpha (kR)}\> .\eqno(15)$$
Upon taking the limit $R \to 0$ for fixed $\alpha$ one readily finds that
$B/A \to 0$ and that
$$f_0^{s=0} \to \left( {1\over 2\pi ki}\right)^{1/2}
\left( e^{-i|\alpha|\pi}-1\right)\eqno(16)$$
which is precisely the usual $m=0$ contribution to the spinless $AB$
scattering amplitude.  As expected, it coincides with (11), the result in
the spin-1/2 case for a repulsive magnetic moment interaction.

On the other hand an expansion of Eqs. (14) and (15) in powers of
$\alpha$ is plagued with logarithmic singularities.  In particular the
lowest nonvanishing contribution gives
$$f_0^{s=0} \cong \left({i\pi\over 2k}\right)^{1/2} \alpha^2 (\ell n
{kR\over 2} + \gamma)$$
where $\gamma$ is Euler's constant.
It is a curious fact that that this happens to agree
 to lowest order with the result of ref. 3
despite the fact that the model considered there is physically quite different.
Also relevant in this context is Nagel's observation that the lowest
order result of ref. 3 is too large by a factor of 3.

\noindent IV.  Conclusion

In this work the perturbative expansion of the AB scattering amplitude
has been considered from the viewpoint of the finite radius flux tube
model.  The latter has successfully been applied in the past to the
problem of calculating exactly the AB amplitude for both spin zero and
spin-1/2 scattering.  Thus it is not surprising that it is also able to
deal with the perturbative approach to this problem.  In the spinor case
the lowest order result of ref. 7 has been confirmed and extended to
arbitrary order in $\alpha$.  For spinless particles it has been shown
that the AB amplitude is an infinite series in $\alpha \ell n kR$ which
can be summed to give the known form of that amplitude.

It is also appropriate to discuss briefly the relevance of this
development to the corresponding problem in Galilean field theory.  The
framework for such an approach was provided in ref. 9 which formulated
the Galilean covariant gauge theory of the Chern-Simons interaction.
Because of the superselection rule on the mass, it was shown there that
one could approach the problem of calculating an arbitrary scattering
process by restricting consideration to an $N$-body sector.  This allowed
one to derive a Schrodinger equation for the $N$ body problem with
each pair interacting as zero radius flux tubes.  Thus the field theory
sector by sector is formally equivalent to obtaining a solution of a
conventional Schrodinger equation.  However, once that set of Schrodinger
equations is obtained, it is necessary to give it a precise meaning by
either stipulating a set of boundary conditions or by regularizing the
interaction.  In the latter case one can invoke the impenetrable solenoid
of ref. 3 or the finite radius flux tube model of ref. 7.  The latter, of
course, is more flexible since it allows for the accommodation of spin.
In view of the equivalence between the field theory and the Schrodinger
equation approaches, it is clear that all the calculations of this paper
apply to both domains.  What would be interesting and worthwhile in this
context would be to write the field theory of ref. 9 {\it ab initio} in
such a way that the methodology of the finite radius flux tube method is
built in at the start.  This can presumably be done by a point splitting
definition of the current operator, but has not yet been carried out.

Another way in which one can approach the field theoretical approach to
the AB scattering calculation is to determine the propagators of the
theory and subsequently use them to carry out perturbative calculations.
This has recently been done by Bergman {\it et al.} [10] who find that for
spinless particles the AB amplitude is logarithmically divergent.  On the
basis of renormalizability considerations they argue for the inclusion of
 a contact interaction
which allows the logarithmic divergence to be eliminated.  Since the
latter type of interaction is formally identical to the $\alpha s 1/r
\delta (r)$ term of the Schrodinger equation for spin-1/2, it is clear
that the cancellation found in ref. 10 to lowest order coincides with the
repulsive (i.e., $\alpha s > 0$) case of the preceding section.  It is
interesting to note, however, that the case $\alpha s < 0$ also leads to
such a cancellation.  This has been shown here in general and can also
be verified using the lowest order calculations of ref. 10.  In that work the
choice of a repulsive contact term was made in order to obtain
agreement in perturbation theory with the AB amplitude.  The opposite
choice is, however, equally allowable and gives not the AB amplitude but
rather an AB-like amplitude in which $|\alpha|$ is simply replaced by its
negative in Eq. (16).

A final comment (which applies to the sector-by-sector approach as well)
has to do with the details of the cancellation of divergences between the
spin (or spinlike) term and the $\alpha^2/r^2$ potential.  These were
shown to cancel unambiguously in the finite radius flux tube model.
However, in the calculations carried out in ref. 10 it is reasonable --
but not compelling -- to identify the cutoffs associated with the two
parts of the interaction.  If their ratio is different than unity, one
must have additional finite terms in the amplitude which destroy the
exact agreement with the perturbative AB amplitude.  Thus in the absence
of a Galilean field theory which is well-defined (i.e., unambiguously
regularized) at the outset, the perturbative approach to AB scattering
must remain only inadequately understood in that case relative to
the corresponding quantum mechanical result.

\noindent Acknowledgements

This work was supported by Dept. of Energy Grant DE-FG02-91ER40685.

\vfil\eject

\noindent References

\item{1.} Y. Aharonov and D. Bohm, Phys. Rev. {\bf 115}, 485 (1959).
\item{2.} E. L. Feinberg, Usp. Fiz. Nauk. {\bf 78}, 53 (1962) [Sov. Phys.
Usp. {\bf 5}, 753 (1963)]; E. Corinaldesi and F. Rafeli, Ann. J. Phys.
{\bf 46}, 1185 (1978).
\item{3.} Y. Aharonov, C. K. Au, E. C. Lerner, and J. Q. Liang, Phys.
Rev. {\bf D29}, 2396 (1984).
\item{4.} B. Nagel, Phys. Rev. {\bf D32}, 3328 (1985).
\item{5.} C. R. Hagen, Phys. Rev. {\bf D41}, 2015 (1990).
\item{6.} F. Vera and I. Schmidt, Phys. Rev. {\bf D42}, 3591 (1990).
\item{7.} C. R. Hagen, Phys. Rev. Lett. {\bf 64}, 503 (1990).
\item{8.} It has been shown in ref. 7 that in the $R=0$ limit the result
is independent of the precise distribution of the magnetic field inside
the tube.  However, a surface distribution of the field is a convenient
assumption since it considerably
simplifies the calculations involved.
\item{9.} C. R. Hagen, Phys. Rev. {\bf D31}, 848 (1985).
\item{10.} O. Bergman and G. Lozano, Ann. Phys. {\bf 229}, 416 (1994);
D. Bak, O. Bergman, Phys. Rev. {\bf D15}, 1994 (1995).

\vfil\eject
\end